\documentclass[
  aps,prl,reprint,groupedaddress,
  amsmath,
  amssymb]{revtex4-2}
\usepackage{graphicx}
\usepackage{dcolumn}% Align table columns on decimal point
\usepackage{bm}% bold math
\usepackage[utf8]{inputenc}
\usepackage[T1]{fontenc}
\usepackage{mathptmx}
\usepackage{etoolbox}
\usepackage{babel}
\usepackage{xcolor}
\usepackage{braket}
\usepackage{parskip}
\usepackage{chngcntr}

\usepackage{import}
\usepackage{xifthen}
\usepackage{pdfpages}
\usepackage{transparent}
\usepackage{hyperref}
\usepackage{orcidlink}

\usepackage{academicons}

\makeatletter
\AtBeginDocument{\let\LS@rot\@undefined}
\def\@email#1#2{%
 \endgroup
 \patchcmd{\titleblock@produce}
  {\frontmatter@RRAPformat}
  {\frontmatter@RRAPformat{\produce@RRAP{*#1\href{mailto:#2}{#2}}}\frontmatter@RRAPformat}
  {}{}
}%
\makeatother

\begin{document}

%\preprint{}

\title[A method for robust spin relaxometry in the presence of imperfect state preparation]{A method for robust spin relaxometry in the presence of imperfect state preparation} %Title of paper
%A method for robust relaxometry in the presence of imperfect state preparation
%Relaxometry in poor polarisation regimes



\author{E.P. Walsh$^*$ \orcidlink{0000-0002-1532-1566}}
%  \altaffiliation[Also at ]{Physics Department, XYZ University.}%Lines break automatically or can be forced with \\
 \email{ewalsh1@student.unimelb.edu.au}
\affiliation{ 
School of Physics, The University of Melbourne, Parkville, VIC 3010 Australia%\\This line break forced with \textbackslash\textbackslash
}%
\affiliation{%
CSIRO Manufacturing, Clayton, VIC 3168, Australia%\\This line break forced% with \\
}%
\author{S. Ahmadi}
\affiliation{CSIRO Manufacturing, Clayton, VIC 3168, Australia%\\This line break forced% with \\
}%

\author{A. J. Healey}
\affiliation{%
School of Science, RMIT University, Melbourne, VIC 3001, Australia%\\This line break forced% with \\
}%

\author{D.A. Simpson$^*$}
 \email{simd@unimelb.edu.au}
\affiliation{School of Physics, The University of Melbourne, Parkville, VIC 3010, Australia%\\This line break forced with \textbackslash\textbackslash
}%

\author{L.T. Hall$^*$}
\email{liam.hall@csiro.au}
\affiliation{CSIRO Manufacturing, Clayton, VIC 3168 Australia%\\This line break forced% with \\
}%

\date{\today}

\begin{abstract}
Spin relaxometry based on quantum spin systems has developed as a valuable tool in medical and condensed matter systems, offering the advantage of operating without the need for external DC or RF fields. Spin relaxometry with nitrogen-vacancy (NV) centers has been applied to paramagnetic sensing using both single crystal diamond and nanodiamond materials. However, these methods often suffer from artifacts and systematic uncertainties, particularly due to imperfect spin state preparation, leading to artificially fast T$_1$ relaxation times. Current analysis techniques fail to adequately account for these issues, limiting the precision of parameter estimation.
In this work, we introduce a minimal fitting procedure that enables more robust parameter estimation in the presence of imperfect spin polarization. Our model improves upon existing approaches by offering more accurate fits and provides a framework for efficiently parallelizing single-spin dynamics studies.
\end{abstract}


\maketitle %\maketitle must follow title, authors, abstract and \pacs

\section{Introduction}
Spin relaxometry is the technique of measuring a nuclear or electron spin decay from a polarized state to thermal equilibrium. 
In recent years, there has been interest in delving into the nanoscale where conventional NMR is limited by stand-off and sensitivity. 
Optically active spins in solid state materials, namely nitrogen-vacancy (NV) centers in diamond, are attractive probes of magnetic noise at this scale, with applications ranging from radical sensing in biological settings to critical phenomena in condensed matter physics. The NV center is desirable for relaxometry owing to its efficient optical state preparation and readout, long relaxation times \cite{balasubramanianUltralongSpinCoherence2009}, photostability, biocompatibility \cite{mohanVivoImagingToxicity2010}, and high sensitivity \cite{hallDetectionNanoscaleElectron2016}. 

Spin relaxometry exploits the NV's relaxation to a thermally mixed state from a polarized one. In the presence of resonant magnetic noise, the intrinsic decay rate to thermal equilibrium accelerates. A change in the intrinsic relaxation rate, in the presence of a paramagnetic target can lead to inferences of amounts of magnetic material \cite{grantNonmonotonicSuperparamagneticBehavior2023, schafer-nolteTrackingTemperatureDependentRelaxation2014, pelliccioneTwoDimensionalNanoscaleImaging2014, sushkovAllOpticalSensingSingleMolecule2014}, chemical structure information \cite{rendlerOpticalImagingLocalized2017,shugayevSynthesisQuantumMetrology2021,peronamartinezNanodiamondRelaxometryBasedDetection2020,abendrothSingleNitrogenVacancyNMRAmineFunctionalized2022, mullerNuclearMagneticResonance2014} and can help identify sources of biomagnetic material \cite{degilleQuantumMagneticImaging2021, cistolaCompactNMRRelaxometry2016,steinertMagneticSpinImaging2013,kaufmannDetectionAtomicSpin2013}. In principle, such a measurement can be highly sensitive and quantitative. However, as we demonstrate in this work, imperfect state preparation can lead to thermal equilibrium being reached more rapidly. The decay of an NV center is typically analyzed using a single exponential \cite{pelliccioneTwoDimensionalNanoscaleImaging2014, sushkovAllOpticalSensingSingleMolecule2014}, returning the decay rate corresponding to the $1/e$ time, $T_1 = 1/\Gamma_1$.  However, such fits cannot account for variations in the state preparation and assume perfect polarization for accuracy. When fitting to an imperfectly polarized NV center it will report on the apparent relaxation rate rather than take any state preparation error into account, limiting sensitivity and producing measurement artifacts in the form of artificially fast intrinsic relaxation rates. Such artifacts manifest in different ways depending on the imaging modality being used. 

Confocal systems are favored for single NV studies due to the spatial resolution that can be achieved as well as efficient state preparation and light collection. Implemented in scanning probes and nanoscale-NMR \cite{schmid-lorchRelaxometryDephasingImaging2015, maletinskyRobustScanningDiamond2012,wrachtrupSingleSpinMagnetic2016}, single, near-surface NV centers offer minimal stand-off from a target and precise knowledge of the intrinsic relaxation rate. However, high laser powers induce unwanted charge state dynamics in near surface NVs\,\cite{bluvsteinIdentifyingMitigatingCharge2019}. To mitigate this, the excitation laser power can be reduced resulting in imperfect polarization and potential measurement artifacts. Confocal studies of NV ensembles can similarly be limited, as the laser powers needed to polarize the entire ensemble are such that charge state changes are likely to occur. 

Widefield imaging is a way of parallelizing spin relaxometry measurements at the cost of axial spatial resolution. A large array of single NVs can be addressed and imaged simultaneously in widefield\,\cite{guoWidefieldFourierMagnetic2024, pfenderSinglespinStochasticOptical2014}. Engineering advances of nanopillars are an emergent platform for such applications owing to their enhanced fluorescence from waveguides leading to improved sensitivity and measurement times\,\cite{mccloskeyEnhancedWidefieldQuantum2020, momenzadehNanoengineeredDiamondWaveguide2015, widmannFabricationCharacterizationSingle2015, neuPhotonicNanostructures111oriented2014, babinecDiamondNanowireSinglephoton2010}. Often, an NV ensemble is preferable owing to improved signal-to-noise and faster integration times while also enabling imaging\,\cite{simpsonMagnetoopticalImagingThin2016, taylorHighsensitivityDiamondMagnetometer2008}. However, in all implementations of widefield relaxometry, uniform and ideal spin polarization is limited: large laser spots result in low power densities and Gaussian beam profiles lead to spatial variation in the power density across a field of view. Poor polarization results in fewer collected photons and ultimately, the number of measurement repetitions required for imaging large areas or building statistics is increased. Longer pulse durations can in principle be used to remedy poor state initialization at the edges of the laser spot, however this comes at the cost of reducing readout contrast in regions that are efficiently initialized.

In this work, we systematically study the effect of imperfect polarization on relaxometry measurements via the dynamics of a single NV center and present an analytic model that we can use to extract the effects of target magnetic noise and experimental constraints. We first apply the model to a simple case of a single NV center initialized with varying laser power and compare fitting fidelity and sensitivity to a single exponential. Second, we consider the case where a distribution of polarization efficiency is recorded simultaneously from an ensemble of NVs using widefield relaxometry.
We show that the model presented herein increases usable field of view and increases sensitivity by maximizing intrinsic relaxation rates.

\begin{figure*}
    \includegraphics[width=\textwidth]{src/figs/FIG1.png}
    \caption{a) Widefield imaging of an ensemble of NV centers is captured with a camera. Laser power can be controlled by changing the linear polarization of the laser through a polarizing beam splitter. Choice of laser power, length of laser application, $t_p$ and beam size all impact the polarization rate, $\Gamma_p$, which limits the polarization efficiency of the NVs, and therefore the PL received. b)i) A spin relaxometry sequence involves laser pulses to polarize the spin state, with increasingly long dark times, $\tau$, for spin relaxation to take place. In widefield (ii), the entire polarization pulse and dark time is typically recorded by the camera due to the length of the camera exposure time. Each dark time and initialization/readout pulse is repeated N times to fill the exposure time. iii) The polarization power limits the amount of polarization achieved, and the recorded amount of relaxation in each $\tau_i$ step. c) An example of a T$_1$ relaxometry curve under different polarization conditions, resulting in different apparent relaxation rates.}
    \label{fig:paper1}
\end{figure*}%\label{}
% \subsection{}
% \subsubsection{}
\section{A two-state model}
NV spin relaxometry is an optical technique, where a green laser is used for state preparation and and readout. Red photoluminescence (PL) is recorded from one or many NVs at once, Figure \ref{fig:paper1}a. The measurement is performed by polarizing spins with an off-resonance green laser pulse for some time, $t_{\textnormal{p}}$, then allowing for some free evolution time, $\tau_i$, before optically reading out the final spin state, Figure \ref{fig:paper1}b,i. The NV spin state can be read out from the PL which can be captured via numerous approaches, including photodiodes and cameras, Figure \ref{fig:paper1}b,ii. Repeating this for a sequence of $\tau_i$  builds up a relaxation curve where, ideally, the $1/e$-time returns the figure of interest, the relaxation rate $\Gamma_1=1/$T$_1$. Limited laser powers and choice of polarizing pulse duration can in some cases result in incomplete state preparation and PL, Figure \ref{fig:paper1}b,iii. In these situations the measured relaxation curve will have a shorter $1/ e$-time and diminished spin contrast, both of which limit sensitivity when trying to observe changes in the relaxation rate Figure \ref{fig:paper1}c. 

\begin{figure}
    \includegraphics[width=\textwidth/2]{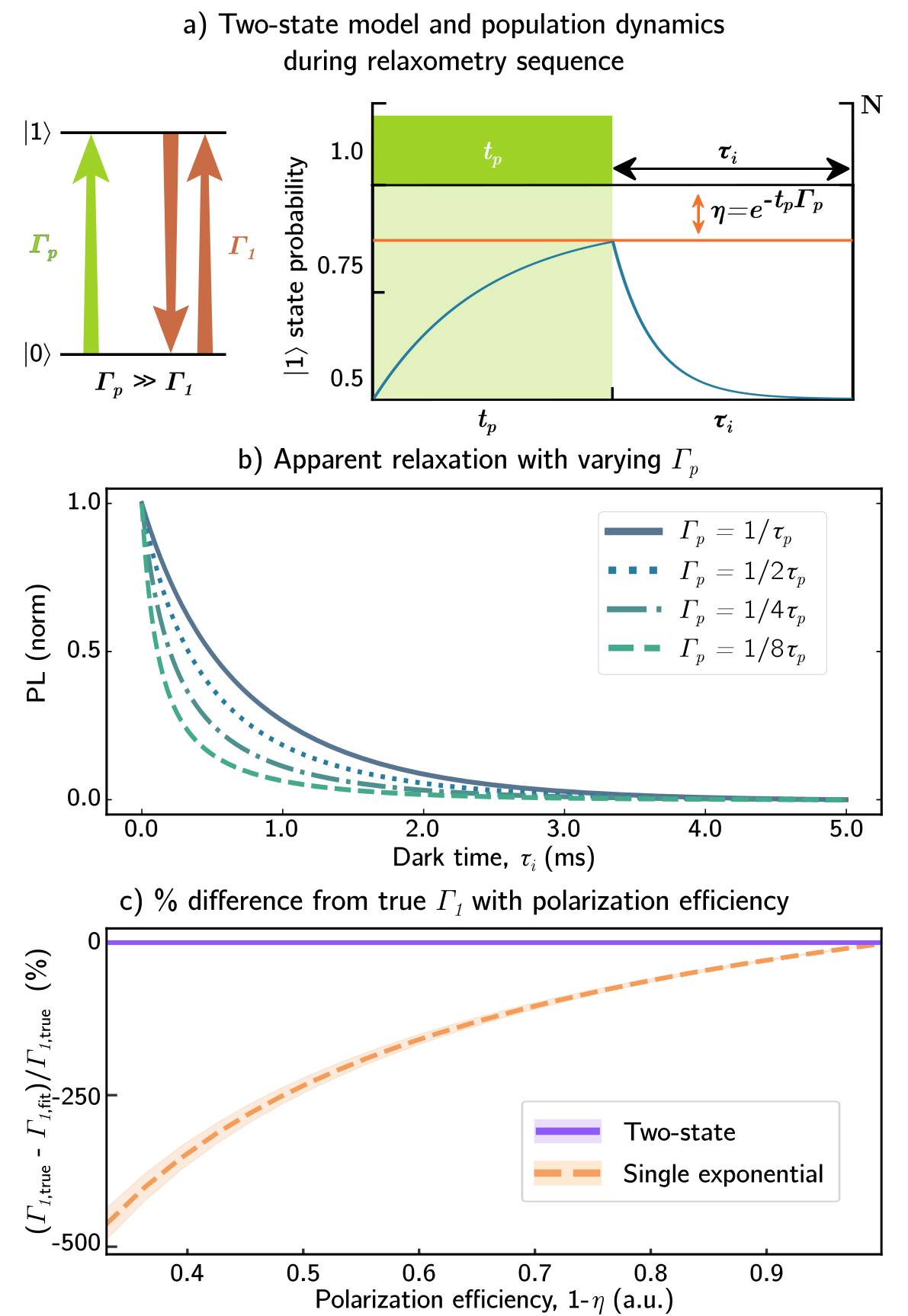}
    \caption{a) The two state system used to model relaxometry decays in the limit that $\Gamma_{\textnormal{p}}\gg\Gamma_1$. A relaxometry sequence is made up of polarization pulses followed by increasingly long dark times. The laser power or pulse duration can both limit the maximum polarization achieved, $\eta$ denotes the effectiveness of the polarization. b) With decreasing polarization rates, $\Gamma_p$, polarization becomes more inefficient and the area under the decay curve can be seen to reduce which results in a shorter $1/ e$-time. c) Even when producing a robust fit to the data, a single exponential will only report the $1/e$-time corresponding to the true relaxation rate if the spin state is perfectly polarized. Otherwise, it will over estimate the relaxation rate as the tail of the curve deviates from a single exponential. The two-state model can account for the imperfect state preparation. \label{fig:paper2}}
\end{figure}

To illustrate this, we propose a two-state model of spin population $\boldsymbol{n} = \displaystyle\left(n_{\ket{0}}, n_{\ket{1}}\right)$, Figure \ref{fig:paper2}a, that enables analytical results and sheds light on measurement artifacts in relaxometry. We will build up the model by replicating experimental procedures used in confocal measurements where time gating is possible and the only source of polarization error is from the excitation laser power pumping rate, $\Gamma_{\textnormal{p}}$ power. Our two state model is then used to analyze the relaxation measured from an ensemble of NVs and target paramagnetic spins using a widefield microscope where spatial effects limit spin polarization. We assume that there are two relevant dynamic processes in the two state model: spin relaxation, $\Gamma_1$, and the pumping rate $\Gamma_{\textnormal{p}}$ which governs the degree of spin polarization. The former, $\Gamma_1$, leads to spin relaxation and is the measurement value of interest, while the latter, $\Gamma_{\textnormal{p}}$, drives the spin into the chosen polarized state, $\boldsymbol{n} \simeq \displaystyle\left(1,0\right)$. 

Given that $\textnormal{T}_1$ relaxometry measurements are incoherent, we do not need to calculate the full dynamics of the system through the Hamiltonian, we need only consider the probabilities of occupying the two states. 
The probabilities are a first order system of coupled linear differential equations, which can be written as a matrix equation,
\begin{equation}
    \frac{d\boldsymbol{n}}{dt} = T\boldsymbol{n} \label{eq:1}, 
\end{equation}
where $T$ is the transition matrix given by the following
\begin{equation}
    T = 
    \begin{bmatrix}
        -\Gamma_1 & \Gamma_1 + \Gamma_\textnormal{p}\\
        \Gamma_1 & -\Gamma_1 -\Gamma_\textnormal{p}
    \end{bmatrix} \label{eq:2}. 
\end{equation}
The solution to Equation \ref{eq:1} is given by,
\begin{equation}
    \boldsymbol{n}(t) = S(t_1\rightarrow t)\boldsymbol{n}(t_1),
\end{equation}
where $S$ is the scattering matrix given by the matrix exponential $S(t)=\exp{(Tt)}$.

In experiments, the polarization and relaxation processes are independent as seen in Figure \ref{fig:paper2}a. Therefore, we are able to separate the two processes into their own propagators $S_{\textnormal{pol}}$ and $S_{\textnormal{relax}}$. During polarization over time, $t_\textnormal{p}$, we can assume the weak relaxation limit, $\Gamma_{\textnormal{p}}\,\gg\,\Gamma_1$, where spin relaxation is negligible during polarization. During spin relaxation, where the state is left to evolve during a dark time, $\tau$, we need only consider the contribution due to $\Gamma_1$ as $\Gamma_{\textnormal{p}}=0$.
\begin{eqnarray}
    S_{\textnormal{pol}} &= 
    \begin{bmatrix}
        1 & 1-\eta\\
        0 & \eta
    \end{bmatrix} \label{eq:7}\;, 
    \\
    S_{\textnormal{relax}} &= \frac{1}{2}
    \begin{bmatrix}
        1+e^{-2\Gamma_1\tau} & 1-e^{-2\Gamma_1\tau}\\
        1-e^{-2\Gamma_1\tau} & 1+e^{-2\Gamma_1\tau}
    \end{bmatrix} \label{eq:8},
\end{eqnarray}
where we define $\eta$, a measure of polarization inefficiency,
\begin{equation}
    \eta = e^{-t_{\textnormal{p}}\Gamma_\textnormal{p}},
\end{equation}
as finite laser duration ($t_{\textnormal{p}}$) and imperfect excitation pumping rate ($\Gamma_{\textnormal{p}}$) will prevent the system from becoming perfectly polarized into the $n_0$ state, Figure \ref{fig:paper2}b. $\eta$ can take values between 0 and 1, where 0 denotes a perfectly polarized system. 

The final state of the system will depend on the number of repetitions, $N$, of the polarization-relaxation cycle. While a single polarization and readout are ideal, experimentally this is impractical, as many repetitions are needed to observe the signal over and above the photon shot noise. Confocal microscope measurements typically use an avalanche photodiode to collect the NV emission at the start of each polarizing laser pulse, repeating the full sweep N times until a desirable signal-to-noise ratio is reached, \ref{fig:paper1}b,ii. In widefield measurement systems, \ref{fig:paper1}b,iii, a camera collects the NV PL, so the exposure time (on the order of tens of milliseconds) dictates a minimum number of repetitions that can occur in a single exposure. 

For simplicity, we assert that the starting population is perfectly polarized $\boldsymbol{n}(0) = \displaystyle\left(1,0\right)$ before each pump and relaxation cycle. Experimentally this is never achieved, however we find that the initial state does not impact the dynamics beyond a few repetitions, so this choice is arbitrary (SI Figure 1). The propagator of the system after $N$ repetitions at each dark time, $\tau_i$, can be found analytically through $S_{\textnormal{eff}} = (S_{\textnormal{pol}}S_{\textnormal{relax}})^N$. The state populations equilibrate after only a few N (SI Figure 1). Experimentally, it is common to have thousands of replicates, hence taking the limit $N\rightarrow\infty$ justifiably allows us to find the analytic solution to the population of as a function of $\eta, \tau$ and $\Gamma_1$.

This gives the population of the polarized state over time, $n_{\ket{0}}(\tau)$, to be,
\begin{equation}
    n_{\ket{0}}(\tau) = \frac{1}{2}\left(1+\frac{-1+\eta}{-e^{2\Gamma_1\tau}+\eta}\right)\label{eq:rel}. 
\end{equation}
We can assert that the apparent relaxation rate is then given by
\begin{equation}
    \Gamma_{\textnormal{app}} = \frac{\Gamma_1}{1-\eta}.
\end{equation}
The effective relaxation rate is what is observed experimentally and approaches the true value as $\eta\rightarrow0$.
Equation \ref{eq:rel} can be generalized to describe the measured, normalized PL and account for baseline noise with a $+c$ offset term,
\begin{equation}
    \textnormal{PL}(\tau) = \frac{\eta-1}{\eta-Ae^{2\Gamma_1\tau}} + c. \label{eq:two_state_expr}
\end{equation} 

To illustrate the utility of this model, we consider a simulation where the pumping rate is reduced relative to the constant pulse duration $t_{\textnormal{p}}$ to keep the $\eta$ unitless. In each case, $\Gamma_{\textnormal{p}}$ is set to decreasing powers relative to the constant pulse duration, $t_{\textnormal{p}}$ to keep $\eta$ unitless. Simulated PL signal is collected over 1000 repetitions of each $\tau$ which range from $1\,\mu$s to $5$\,ms and the relaxation rate, $\Gamma_1$ (T$_1$) is set to 0.5\,kHz (2\,ms). Generated decay curves in Figure \ref{fig:paper2}b show that when each curve is normalized, the apparent $1/e$-time is significantly changed. Fits using a single exponential and the two-state model, Equation \ref{eq:two_state_expr},  were performed on the simulated data using the \texttt{lmfit} Python package. As shown in Figure \ref{fig:paper2}c, fitting a single exponential to an imperfectly polarized system will result in measurement artifacts with misreported $\Gamma_1$. With perfect state preparation ($\eta=0$), the curve is a true single exponential, as expected. Though the single exponential continues to produce well performing fits to the data at low $\eta$ ($\eta<0.4$) (SI Figure 2), the fitted $\Gamma_1$ in this case is overestimated by $\sim50\%$ at the point of $\eta\approx0.4$. As the $1/e$-time decreases with loss of pumping rate, it becomes harder to fit with a single exponential. This is not merely an artifact of an all-optical relaxometry sequence and cannot be accounted for by using $\pi$-pulses to normalize out charge and laser fluctuations, it is a distinct decay rate that obscures the true, intrinsic $\Gamma_1$. In contrast, the two-state model can account for the changes in in the shape of the relaxation curve by characterizing $\eta$ in its fit. This results in recovery of the true $\Gamma_1$ rate in the presence of imperfect spin polarization as shown in Figure \ref{fig:paper2}c. 

In the case of the NV center, the intrinsic relaxation rate, $\Gamma_1$, can also accelerate in the presence of paramagnetic targets with a fluctuation spectrum that overlaps with the NV transition frequency, on the order of GHz \cite{degenQuantumSensing2017, hallDetectionNanoscaleElectron2016, woodWidebandNanoscaleMagnetic2016}.  
In experiments, a characterization measurement of the inherent relaxation rate ($\Gamma_{\textnormal{intrinsic}}=\Gamma_1$) is taken before the introduction of an paramagnetic target followed by another relaxometry sequence. The difference in measured $\Gamma_1$ rates returns the decay rate attributed to the external paramagnetic target,
\begin{equation}
    \Delta\Gamma= \Gamma_{\textnormal{measured}}- \Gamma_{\textnormal{intrinsic}} = \Gamma_{\textnormal{target}}.\label{eq:target_rate}
\end{equation}
Any errors in fitting to such data will compound, leading to inaccurate inferences of target quantities/concentration. Additionally, a slow intrinsic relaxation rate is desirable for maximizing the sensing range and limit of detection. The over estimation of $\Gamma_1$ by a single exponential in low polarization regimes will not only induce measurement errors but also limit the range of detectable signals.

\section{Experiment}
In widefield studies of NVs, there are additional considerations to a relaxometry experiment beyond polarizing power. First, regardless of choice of laser power, there will always be spatial variation in pumping rate $\Gamma_{\textnormal{p}}$ across the laser spot as seen in Figure \ref{fig:paper3}a. Typically,  we will choose to only image a small area of maximum polarization to focus on the most polarized NV centers. Assuming measurements are shot-noise-limited, another limitation is that in the majority of widefield demonstrations the detection cameras used are not able to gate to collect NV emission over the desired timescale. Consequently, laser pulse lengths, $t_{\textnormal{p}}$, are shortened to maximize spin contrast. In most studies of ensemble NV relaxometry, a stretched exponential is able to fit with low error to the relaxation curve measured, however the fitted effective relaxation rate is not well defined with respect to the underlying distribution of rates \cite{johnstonStretchedExponentialRelaxation2006} (SI Figure 3). The two-state model, though built upon a single NV structure is well suited to fitting to the mean of the relaxation rates within the bulk when the distribution can be approximated by a Gaussian (see SI). To demonstrate the impact of imperfect polarization in widefield, we compare the induced relaxation rate from identical paramagnetic targets across a field of view when analyzed using the conventional, stretched exponential and the two-state model.

The NV doped diamond sample used in this study was a $2$\,mm\,$\times$\,$2$\,mm commercially available (Element 6) CVD, $[100]$-oriented diamond was implanted at 4\,keV with a fluence of 10$^{13}$\,cm$^2$ and annealed at 900\,C. Given typical N-to-NV conversions at this energy are around 1\%, we assume an NV density of $1$\,ppm \cite{tetienneSpinPropertiesDense2018a, pezzagnaCreationEfficiencyNitrogenvacancy2010}. 

Experimentally, a custom-built inverted widefield microscope (see SI) with a laser power density of 14 kW/$\mu$m$^2$ over a laser spot with $100$\,$\mu$m diameter. A low ($\sim30$\,G) magnetic field was aligned along one of the four NV axes to isolate a single NV family on which to perform the measurement. During the relaxometry sequence, a 10\,$\mu$s initialization and readout pulse, $t_{\textnormal{p}}$, was used with a 30\,ms exposure time on the camera. Dark times were spaced exponentially from 10\,$\mu$s to 9\,ms. The relaxometry sequence was integrated for $\sim12$ hours to achieve an optimal signal-to-noise ratio on a single pixel level. Normalization was performed by repeating the T$_1$ sequence using a $\pi$-pulse of 300\,ns to invert the spin state. This is done for common mode rejection to account for laser fluctuations\,\cite{jarmolaTemperatureMagneticFieldDependentLongitudinal2012}. 

As is typical for such an experiment, a reference measurement of the bare diamond was taken first, then a $2\,\mu$L droplet of paramagnetic microparticles (Spherotech, AMS-40-10H) suspended in MilliQ at a concentration of $100\,\mu$g$/$mL were deposited on the diamond. The MilliQ was left to evaporate, settling the microparticles on the diamond surface. The relaxometry sequence was then repeated for the same length of time. The region analyzed, Figure \ref{fig:paper3}a, was cropped from the full field of view (FOV) of $512\times512$ pixels to focus on a region with well dispersed microparticles across the full range of pump laser powers, including a region with very little illumination, as seen in the bottom left of Figure \ref{fig:paper3}a. The total region analyzed of the relaxation image was $362\times362$ pixels without binning.

\section{Analysis}
\begin{figure}
    \includegraphics[width=\textwidth/2]{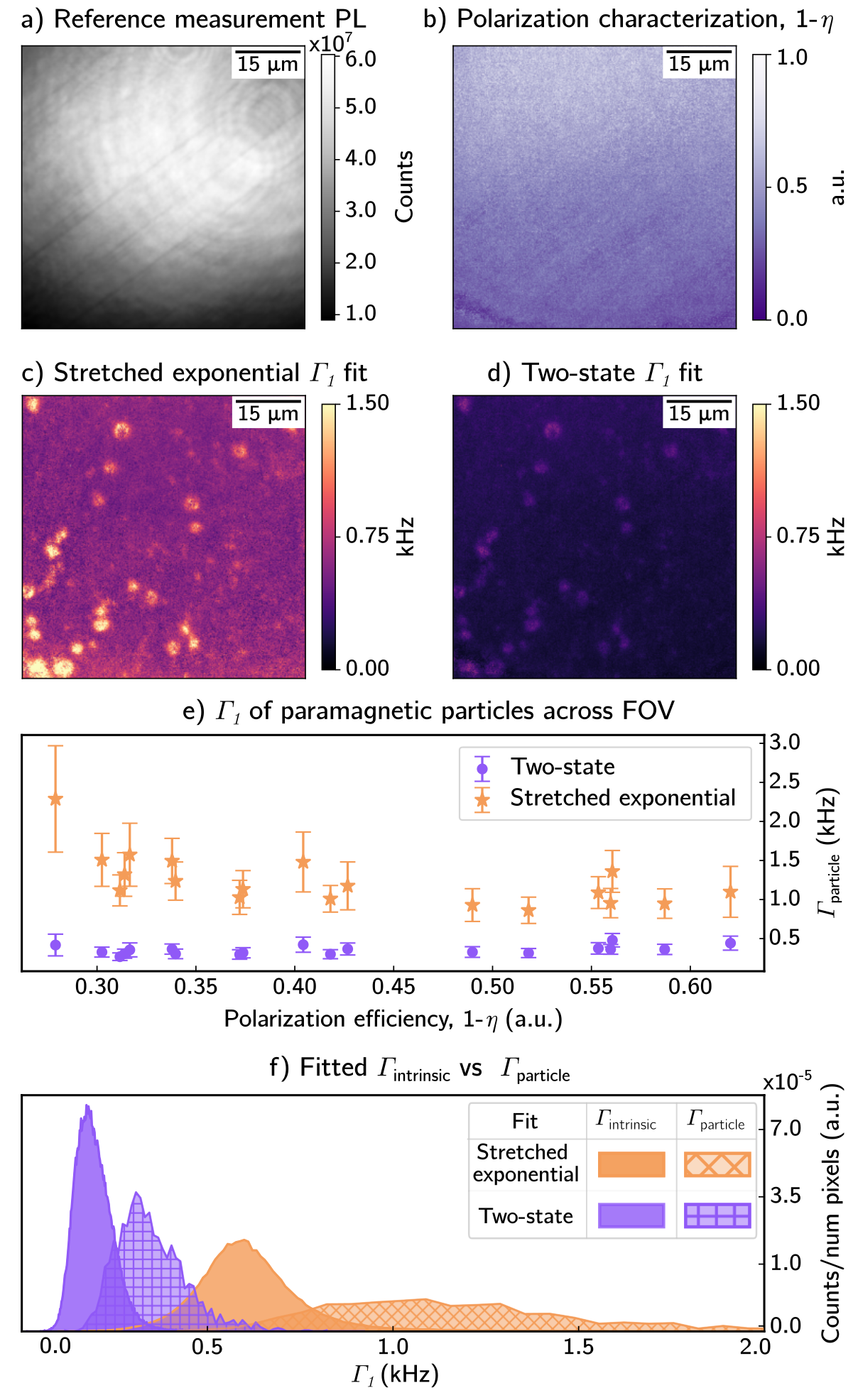}
    \caption{a) The PL from the reference measurement of the bare diamond, used to characterize $\eta$ (b). The characterization map is then used to inform the fit to the relaxometry measurement of the paramagnetic particles. The map of the relaxation rates in the presence of the particles is shown for the (c) stretched exponential and two-state model (d) fits. e) The average fitted relaxation rates of the particles becomes highly variable for low ($1-\eta < 0.35$) polarization regimes when using a stretched exponential. The two-state model remains consistent across the FOV. f) Histograms of the fits to particles and the background on a single pixel level reveal the underlying distribution and uncertainty in stretched exponential fits in a widefield setting. The two-state model increases rate estimation accuracy with smaller standard deviations and improves sensitivity a with slower intrinsic $\Gamma_1$.\label{fig:paper3}}
\end{figure}

To show the utility of our method we now demonstrate an example implementation. In a widefield experiment, a distribution of $\eta$ values will be generally correlated with a laser spot, Figure \ref{fig:paper3}a. We can estimate $\eta$ by asserting that at each pixel, the true relaxation rate should be that phonon-limited \cite{redmanSpinDynamicsElectronic1991, walkerT5SpinLattice1968} and sample dependent noise based on nitrogen and vacancy cluster densities \cite{jarmolaTemperatureMagneticFieldDependentLongitudinal2012}. 
Where a single exponential can be fit to the data we can assume that a true relaxation rate has been measured such as shown by Jarmola \textit{et al}.  \cite{jarmolaTemperatureMagneticFieldDependentLongitudinal2012}, and that the environment surrounding each NV is assumed to be uniform. Jarmola \textit{et al}.  \cite{jarmolaTemperatureMagneticFieldDependentLongitudinal2012} arrive at an empirical formula for an ensemble relaxation rate at room temperature:
\begin{equation}
    \Gamma_1 = A_1 + 175.5\,\textnormal{s}^{-1},
\end{equation}
where $A_1$ is linearly proportional to the NV density of the sample with a gradient of $0.82$\,s$^{-1}$ppm$^{-1}$. For our sample, this would give $A_1 = 0.82$\,s$^{-1}$, resulting in an optimal relaxation rate of $\sim180$\,Hz. Using this ideal relaxation rate, the reference measurement of bare diamond was fitted using the two-state model with $\eta$ as a free parameter and $\Gamma_1$ fixed. Following this procedure we arrive at an polarization characterization ($1-\eta$) map Figure \ref{fig:paper3}b which we can insert into subsequent analysis. For this experimental set up, the maximal polarization region had an average of $\eta \simeq 0.36$ while the regions of minimal polarization correspond with the edge of the laser spot where $\eta \simeq 0.77$. Characterizing $\eta$ this way assumes that there is no additional polarization effects owing to nitrogen and/or vacancy densities or the the paramagnetic targets themselves\,\cite{t.flinnNitrogenVacancyDefects2023} and that there are no changes in laser power between measurements. 

Additionally, we note that Equation \ref{eq:target_rate} ignores the impact of surface noise that is likely relevant in our shallow ensemble\,\cite{myersProbingSurfaceNoise2014}, and thus represents a lower bound on our true ensemble $\Gamma_1$ and therefore $\eta$. In the SI we instead consider a lower value of $333$\,Hz for the intrinsic $\Gamma_1$ (T$_1=3$\,ms), which gives a maximal polarization efficiency of 1 and shows that our model performs similarly. In practice other strategies for precise $\eta$ calculation include confocally probing for the relaxation rate at a under a perfectly polarized state, when $\eta=0$\,\cite{robledoSpinDynamicsOptical2011}. However, in an ensemble this requires large laser powers and the risk of charge state changes is much larger. The method presented here, at the cost of exact rate estimation, only requires access to a widefield microscope.

We next drop cast the paramagnetic microparticles onto the surface of the diamond and the relaxometry measurement was repeated. Figure \ref{fig:paper3}c shows the relaxation rate from a stretched exponential fit across the FOV. The values obtained across the image are variable and correlated with the laser spot. There is also a strong covariance with the stretch parameter (see SI) which contributes to the uncertainty in values at individual pixels. Figure \ref{fig:paper3}d shows the two-state fit obtained after inserting the characterized $\eta$ values into the same data set. Values are much more consistent across the FOV. The characterization of $\eta$ is an essential step owing to the correlation between $\eta$ and $\Gamma_1$ in Equation \ref{eq:two_state_expr}. Figure \ref{fig:paper3}e compares the relaxation rates measured for individual paramagnetic particles across the FOV for the two analysis methods. An average $\Gamma_1$ value for each particle was calculated from a $10\times10$ pixel region centered on the particle (SI Figure 4), and error bars are indicative of the standard deviation over this region. We can see that the two-state fit produces consistent values across the full range of polarization efficiencies, while the stretched exponential data points are more variable and deviate strongly for polarization efficiencies $<0.45$. In practice this restricts the useful field of view to $\sim 50\,\mu$m for the stretched exponential fit while the two-state fit is robust over the full FOV. Finally, a histogram of the single pixel fits further reveals the advantages of the two-state model, Figure \ref{fig:paper3}f. The background ($\Gamma_{\textnormal{intrinsic}}$) and target ($\Gamma_{\textnormal{particle}}$) rate histograms were sampled from the entire image. The pixels that were not analyzed as targets in Figure \ref{fig:paper3}e were all used for the $\Gamma_{\textnormal{intrinsic}}$ histograms. The histograms were then normalized to the total number of pixels. 

The background relaxation rate fitted by the stretched exponential had a mean and standard deviation of $0.626 \pm 0.177$\,kHz while the two-state model returned $0.197 \pm 0.052$\,kHz.  The difference in two-state average rate from the characterization rate of $180$\,kHz can also be explained by slight differences in focus of the laser between the reference measurement and that with the particles, resulting in slight variation in polarization of the ensemble between measurements. Despite this, the stretched exponential fits to a faster baseline rate and a broader distribution than the two-state model. Additionally, when comparing the averages of the paramagnetic target histograms, the stretched exponential and two-state fits reported $\Gamma_{\textnormal{particle}} = 1.242 \pm 0.444$\,kHz and $0.353 \pm 0.093$\,kHz respectively. The two-state model is able to fit convincingly to an average induced relaxation rate with minimal artifacts from the polarization variation across the sample while the its impact on the stretched exponential is much more pronounced.

While the exact relaxation rates that are fit using the two-state method are dependent on your characterized $\eta$ values, the homogeneity in $\Gamma_1$ across the FOV is always gained over a stretched exponential.

\section{Conclusion}
It has been shown that in experiments where perfect polarization cannot be reached, single and stretched exponentials are inappropriate expressions for fitting relaxometry decays as they introduce measurement artifacts. We proposed an alternate model for single NV center decays that accounts for the effects of imperfect state preparation. As a result, workable fields of view can be increased and target rate estimation can be more specific. Additionally, relaxometry of near-surface NVs where charge state stability is a concern can now be performed with greater sensitivity and accuracy. These improvements provide a pathway for efficiently parallelizing single-spin dynamics studies such as in nanopillar arrays. The model shown here is one possible implementation and an interesting avenue for future work could be to consider other options for characterizing the polarization, potentially integrating MW inhomogeneity to enable using ODMR or Rabi contrasts to characterize $\eta$ and save time by avoiding a reference relaxometry measurement. Additionally, exploring other ways of constraining $\eta$ from PL information could allow for both $\eta$ and $\Gamma_1$ to be free parameters in a fit.

\section{Acknowledgments}
The research was supported by the Australian Research Council Centre of Excellence in Quantum Biotechnology through project number CE230100021 and the Australian Research Council Centre of Excellence in Exciton Science through project number CE170100026. E.P.W. is supported by an Australian Government Research Training Program
Scholarship. L.T.H. acknowledges the support of the Commonwealth Scientific and Industrial Research Organisation via a ResearchPlus Science Leader Grant. D.A.S would like to acknowledge support from the ARC Mid-Career Industry Future Fellowship (IM240100073) and the MCN Technology Fellow Ambassador Program.

\section{Data availability}
The data presented in the work can be provided from the corresponding authors at request.

\appendix
% \counterwithin{figure}{section}
\section{Supplementary information}
\subsection{Further mathematical details of the model \label{sec:SI1}}
Equations 1 and 2 lead to the following matrix equation,
\begin{equation}
    \frac{d\boldsymbol{n}}{dt} = 
    \begin{bmatrix}
    -\Gamma_1 n_0(t)+\Gamma_1 n_1(t)+\Gamma_p n_1(t)\\
    \Gamma_1 n_0(t)-\Gamma_1 n_1(t)-\Gamma_p n_1(t)
    \end{bmatrix}.
\end{equation}
The matrix exponential can be expressed as follows, 
\begin{equation}
    S(t) = \exp{(Tt)}
\end{equation}
\begin{equation}
    = 
    \begin{bmatrix}
        \frac{\Gamma_1+\Gamma_{\textnormal{p}}}{2 \Gamma_1+\Gamma_{\textnormal{p}}}+\frac{\Gamma_1 e^{t (-2
        \Gamma_1-\Gamma_{\textnormal{p}})}}{2 \Gamma_1+\Gamma_{\textnormal{p}}} & \frac{\text{$\Gamma
        $1}+\Gamma_{\textnormal{p}}}{2 \Gamma_1+\Gamma_{\textnormal{p}}}-\frac{(\Gamma_1+\Gamma_{\textnormal{p}}) e^{t (-2
        \Gamma_1-\Gamma_{\textnormal{p}})}}{2 \Gamma_1+\Gamma_{\textnormal{p}}} \\
        \frac{\Gamma_1}{2 \Gamma_1+\Gamma_{\textnormal{p}}}-\frac{\Gamma_1 e^{t (-2 \Gamma_1-\Gamma_{\textnormal{p}})}}{2 \Gamma_1+\Gamma_{\textnormal{p}}} & \frac{\Gamma_1}{2 \Gamma_1+\Gamma_{\textnormal{p}}}-\frac{(-\Gamma_1-\Gamma_{\textnormal{p}}) e^{t (-2 \Gamma_1-\Gamma_{\textnormal{p}})}}{2 \Gamma_1+\Gamma_{\textnormal{p}}}
    \end{bmatrix}\label{eq:6}. 
\end{equation}

The state of the system at any time, and for arbitrary pulse repetitions is given by $n(\tau) = S_{\textnormal{eff}}\,n(\tau_0)$,
\begin{equation}
    \boldsymbol{n}(\tau) = 
    \begin{bmatrix}
    \frac{e^{2 \Gamma_1 \tau }-2 \eta +\eta ^N e^{-2 \Gamma_1 (N-1) \tau }-\eta ^N e^{-2\Gamma_1 N \tau }+1}{2 \left(e^{2 \Gamma_1 \tau }-\eta \right)} \\
    -\frac{\left(e^{2 \Gamma_1 \tau }-1\right) \left(\left(\eta  e^{-2 \Gamma_1 \tau }\right)^N-1\right)}{2 \left(e^{2 \Gamma_1 \tau }-\eta \right)}
    \end{bmatrix}\label{eq:pop2state}.
\end{equation}
If we fix the polarization time, $t_{\textnormal{p}} = \frac{1}{2\Gamma_p}$, such that it is only being driven for half of it's $1/e$ time to polarization, then we have a polarization error of $\eta \approx 0.61$. The effect of $N$ on the probability of occupying the $|0\rangle$ state can be seen in SI Figure \ref{fig:varyN}a where the steady state is not immediately reached. This can be characterized through the comparison of successive changes in area under each graph, SI Figure \ref{fig:varyN}b) which shows how quickly the state is able to converge with respect to varying polarization errors, $\eta$. As $\eta$ is reduced (better polarization), the rate of convergence to the steady state increases. 
\renewcommand{\thefigure}{SI \arabic{figure}}
\setcounter{figure}{0}
\begin{figure*}[h]
    \includegraphics[width=\textwidth]{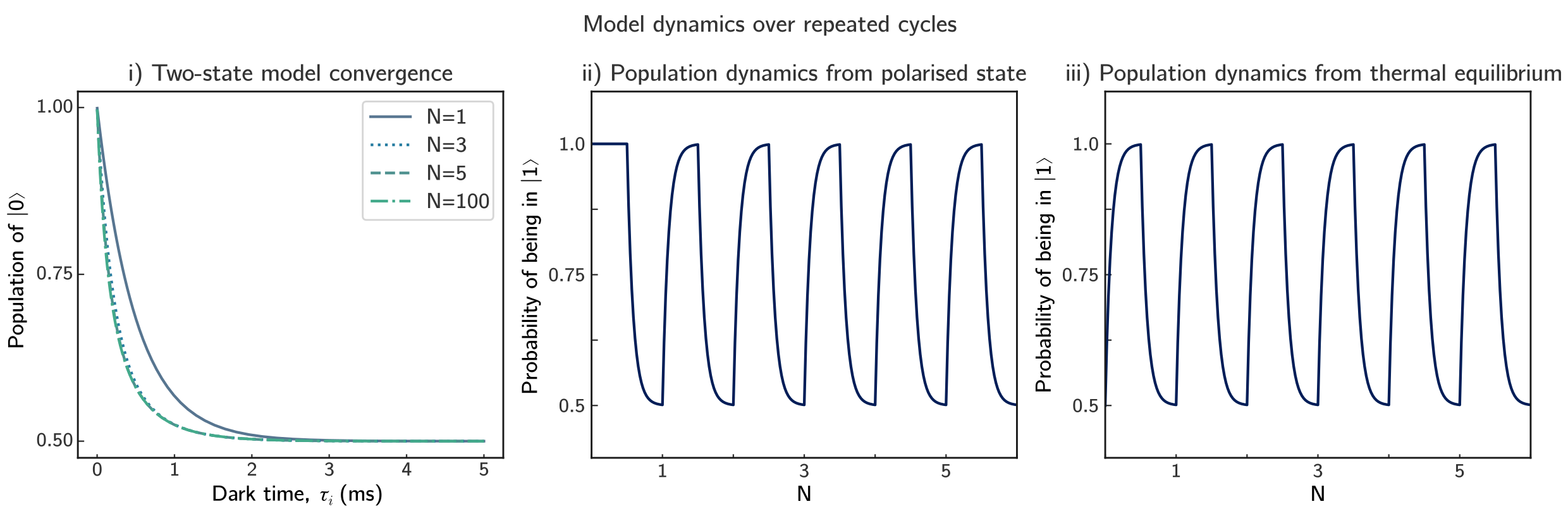}
    \caption{a) Probability of occupying $|0\rangle$ can be seen to converge as N increases. b), c) Population of initial state plotted as a function of time through the sequence in Figure 2a as an analogy to PL measured in experiment with varying polarization time. \label{fig:varyN}}
\end{figure*}

SI Figure \ref{fig:varyN}a) shows how the two state dynamics converge to the analytic solution after only a small number of N repetitions. This indicates that this model is still valid in low-repetition regimes as polarization equilibrium is reached within a few pulses for this system. Additionally, with respect to mimicking experimental procedure by asserting that the initial population be fully polarized before each repetition, the model indicates that the initial polarization condition has no long term impact on the ultimate populations of the states beyond $\approx5$ repetitions during the duty cycle, SI Figure \ref{fig:varyN}b. 
\subsection{Fit details for the single exponential case}
A subset of fits to the simulated confocal example are shown here. In high polarization regimes, the single exponential looks to provide a good fit to the data however, as the $1/e$-time reduces as a result of poor polarization a single exponential no longer has the flexibility to account for this.
\begin{figure*}[h!]
    \includegraphics[width=\textwidth]{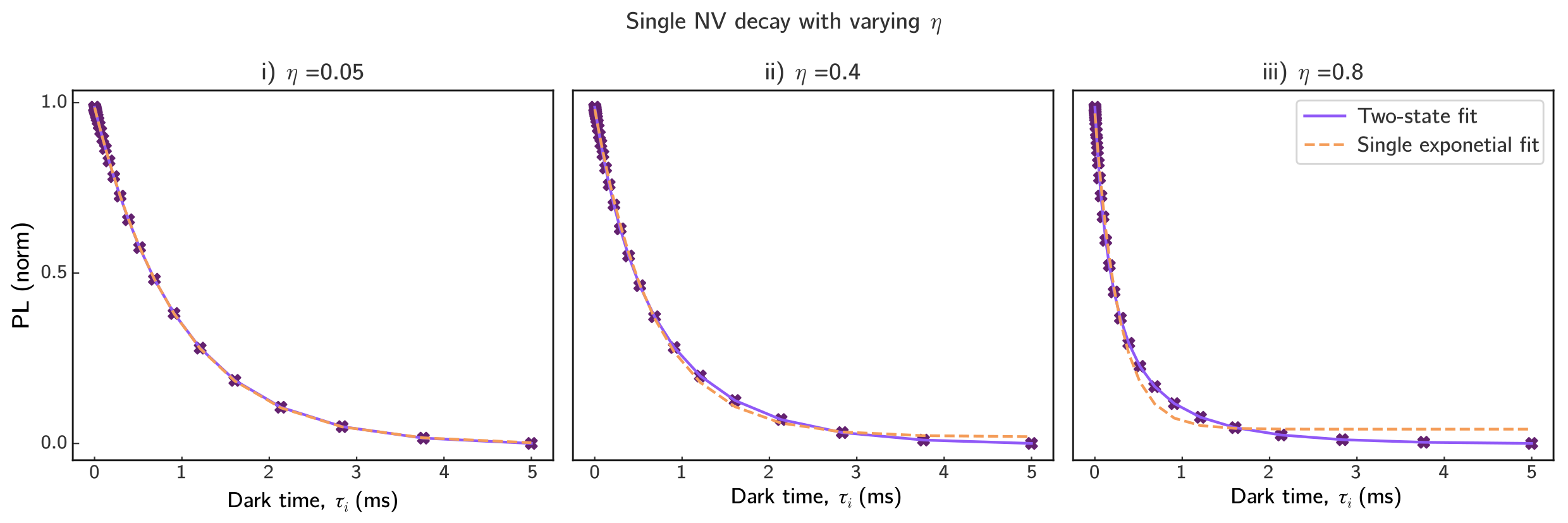}
    \caption{Single exponential fitting of relaxation curves under varying polarization, $\eta$. \label{fig:sec2fits}}
\end{figure*}

\subsection{Extension to an ensemble of NV centres}
\subsubsection{Custom widefield microscope details}
A 532\,nm excitation source (2 W, Opus 532, Laser Quantum) was used. The diamond was mounted with UV glue (Norland Optical Adhesive 63) on to a gold MW antenna evaporated onto a glass coverslip to enable a reference measurement, MWs were generated using an Agilent (N5183A MXG) Analogue Signal Generator and amplified with a Mini-Circuits amplifier (HPA-50W-63+). Imaging was performed with a 40x oil objective (1.3 NA, Plan Fluor, Nikon) resulting in a maximized power density of 14\,kW/$\mu$m$^2$. The PL was filtered from the excitation light through a dichroic filter (Semrock, FF605-Di02-25x36) and subsequent $550$ and $600$\,nm long-pass filters (Thorlabs FELH0550, FELH0600) before being focused onto a scientific CMOS camera (Andor Zyla). 

\subsubsection{Fitting to an ensemble with the two-state model}
A stretched exponential is defined by,
\begin{equation}
    PL(\tau) = Ae^{-(\tau\Gamma_1)^p}+c,
\end{equation}
where $p$ is the stretch parameter. 
As discussed in the main manuscript, a stretched exponential does not report on the average rate of an ensemble. While a stretched exponential can be fit the the relaxation curve, Figure 3a (i-iii), for a wide range of $\eta$, we show how the reported relaxation rate from the stretched exponential increases with decreased spin polarization. The two state model, however, reports on the mean of the underlying distribution.
\begin{figure*}[h!]
    \includegraphics[width=\textwidth]{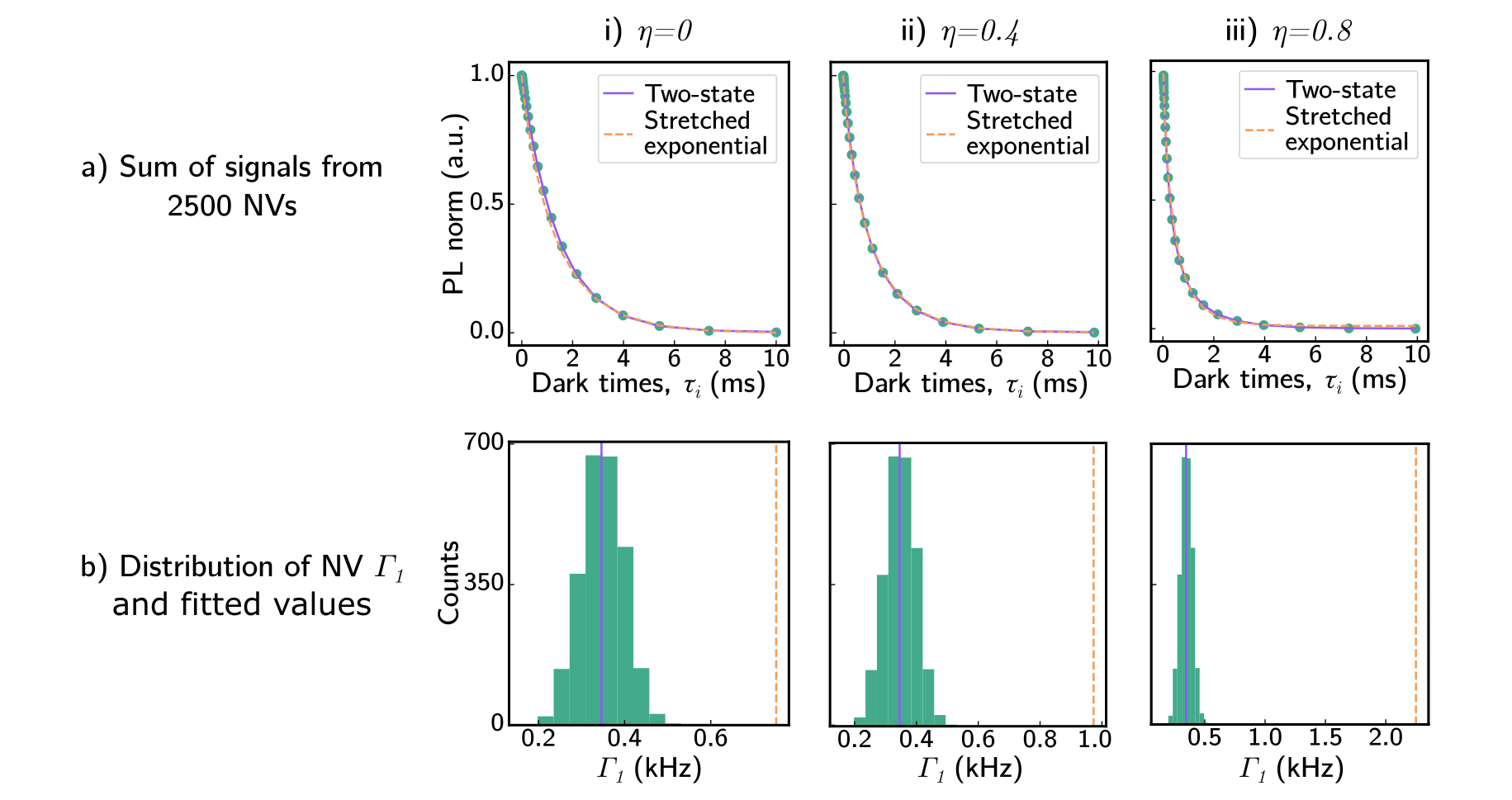}
    \caption{A comparison of fits (a) to an ensemble of NVs with rates sampled from a normal distribution (b). Even when the entire ensemble is perfectly polarized (i), a stretched exponential does not fit to the average of the underlying distribution, unlike the two state model which stays robust over the full range of polarization efficiencies (ii, iii). Furthermore, as the ensemble is less efficiently polarized, the stretched exponential returns faster rates.\label{fig:two_state_fig_params}}
\end{figure*}

\subsection{Analysis of paramagnetic microparticles}
The paramagnetic microparticles were isolated in $10\times10$ pixel regions as shown in SI Figure 4a. The histograms presented in Figure 3f of the main manuscript were sampled from the isolated pixel regions for $\Gamma_{\textnormal{particle}}$ ($\Gamma_{\textnormal{intrinsic}}$) in SI Figure 4b (4c). Blacked out regions were not included in the respective histograms.

\begin{figure*}[h!]
    \includegraphics[width=\textwidth]{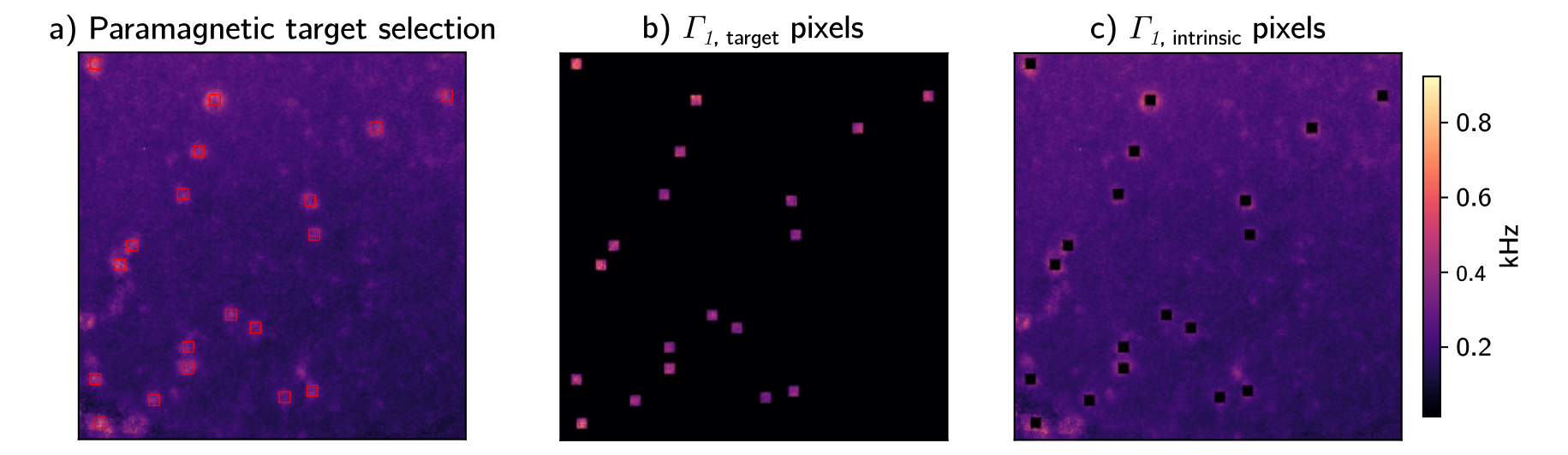}
    \caption{a) In total 19 particles were analyzed, shown by the red squares. b) The $10\times10$ pixel regions that contributed to the histogram of $\Gamma_{1, \textnormal{target}}$. c) The pixels that remained after (b) contributed to the histogram of $\Gamma_{1, \textnormal{intrinsic}}$. \label{fig:single_pixel_analysis}}
\end{figure*}
In the two figures below, images of the respective two-state model and stretched exponential fit parameters are shown. 
\begin{figure*}[h!]
    \includegraphics[width=\textwidth]{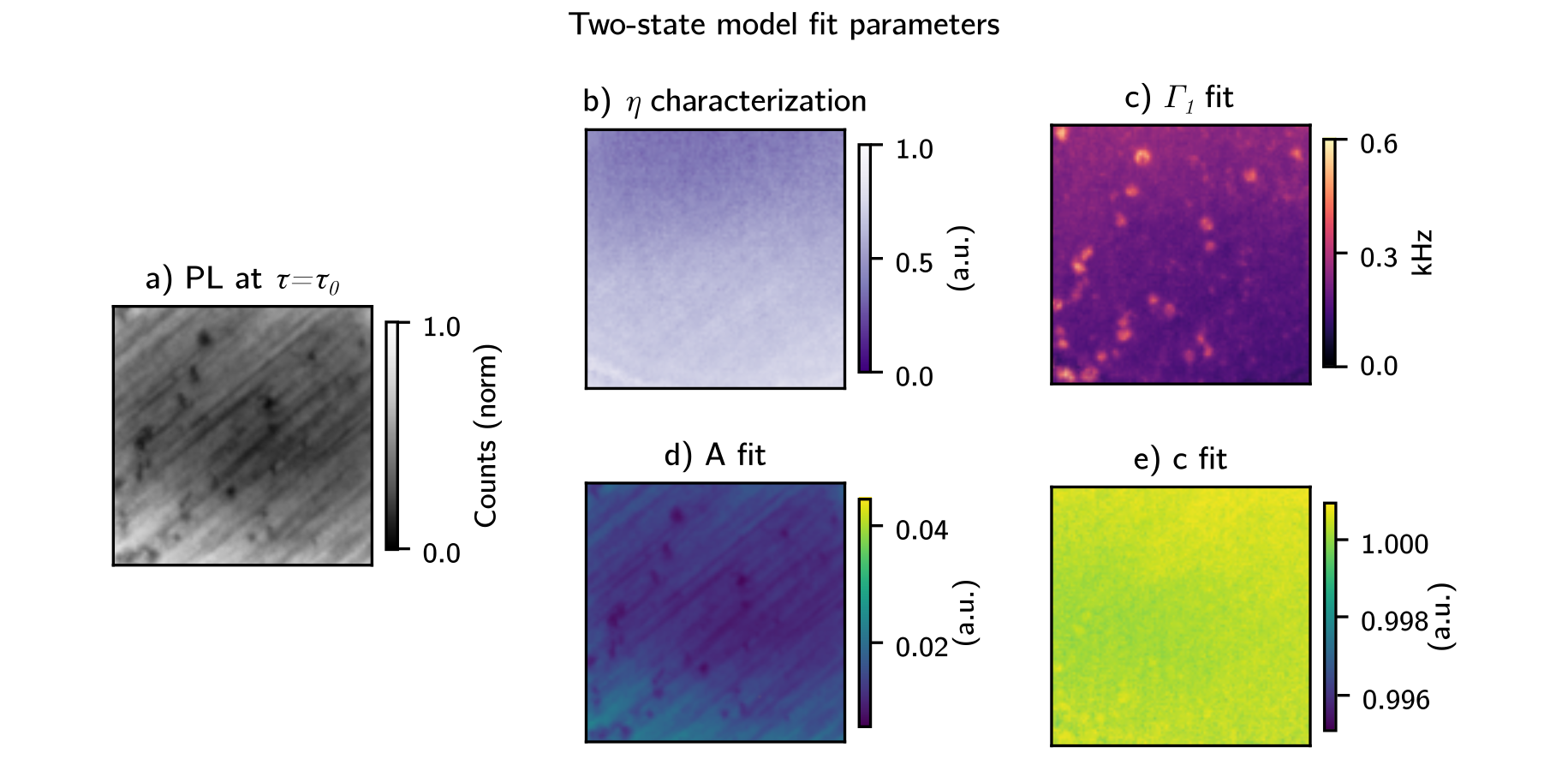}
    \caption{Fit parameters per pixel from the two-state model  (Equation 9) in the manuscript. \label{fig:two_state_fig_params}}
\end{figure*}

\begin{figure*}[h!]
    \includegraphics[width=\textwidth]{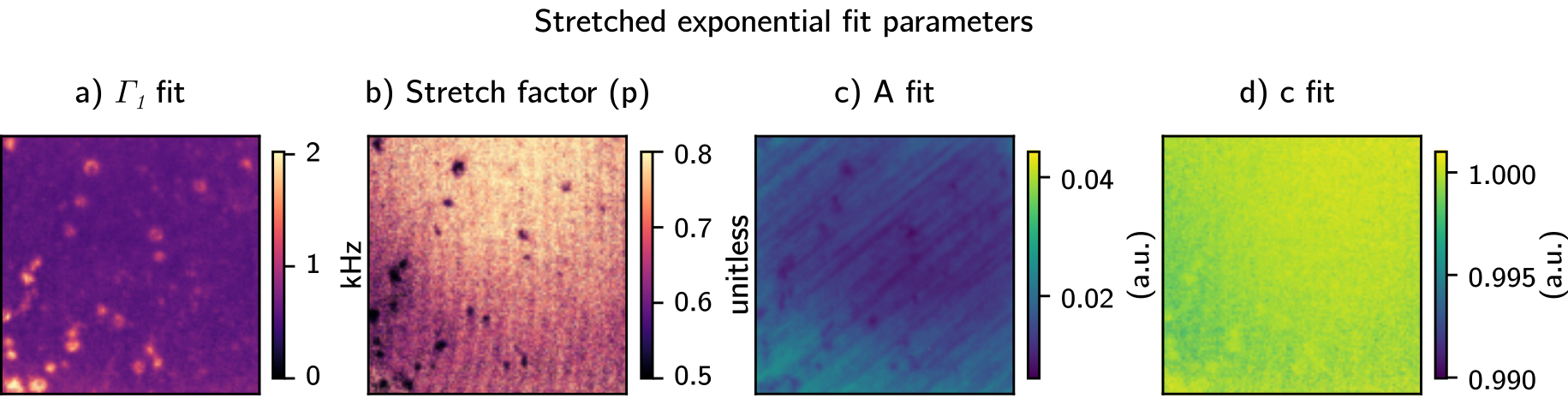}
    \caption{Fit parameters per pixel from the stretched exponential in the manuscript. \label{fig:stretch_fit_params}}
\end{figure*}

\begin{figure*}[h!]
    \includegraphics[height=15cm]{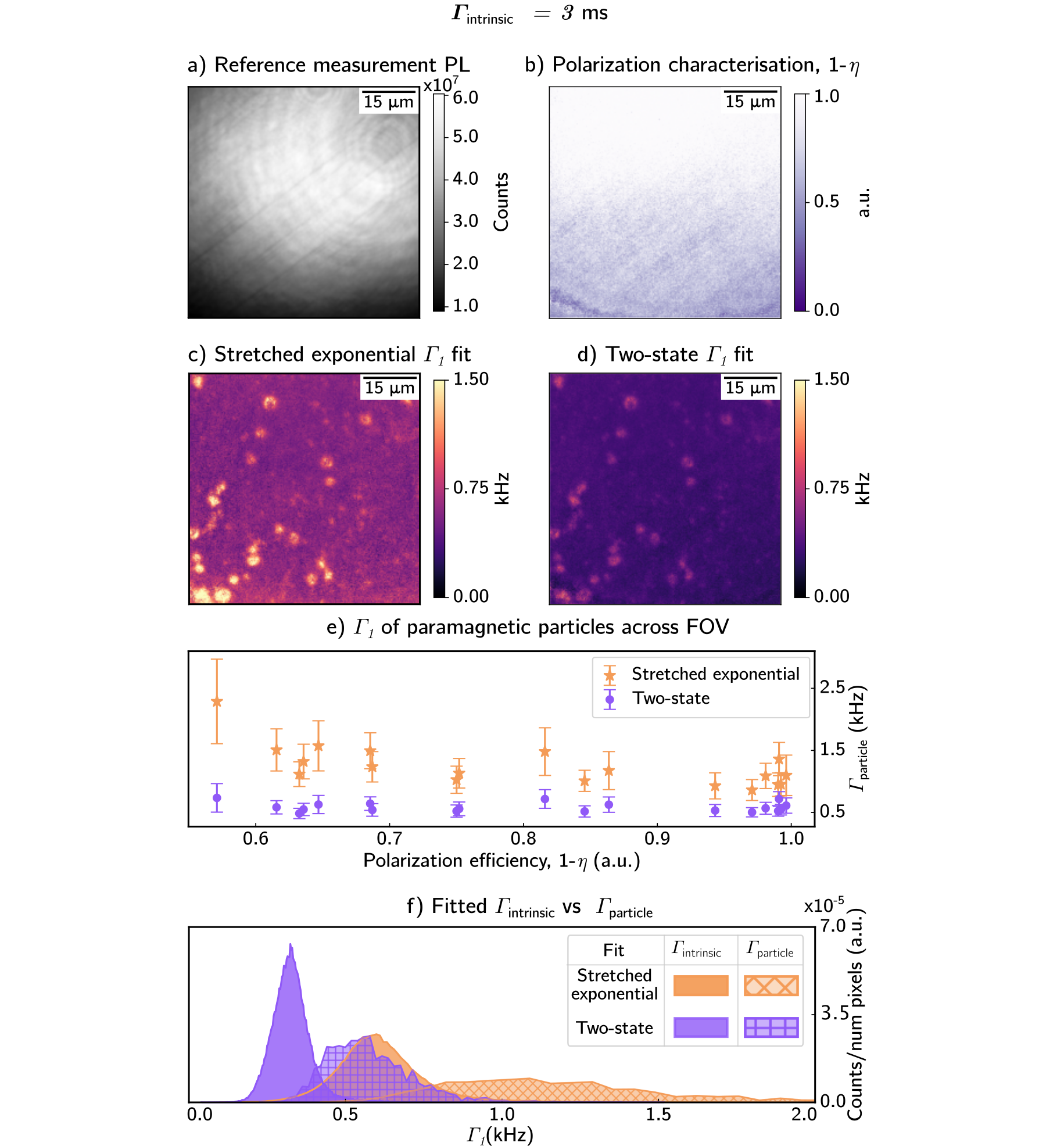}
    \caption{When constraining the maximum $\Gamma_1$ to a less optimal rate, the two-state fit remains more uniform across the FOV compared to a stretched exponential fit and the rates on the particle regions are more consistent. The two-state returns a background $\Gamma_{\textnormal{intrinsic}}$ of $0.331 \pm 0.067$\,kHz and $\Gamma_{\textnormal{particle}}$ of $0.585 \pm 0.138$\,kHz. \label{fig:eta_maximised}}
\end{figure*}

\bibliography{src/betterlib}
% \addbibresource{src/reflib.bib}

\end{document}